\documentclass[11pt]{article}

\usepackage{arxiv}
\usepackage{amssymb,amsmath,amsthm}
\newtheorem{theorem}{Theorem}
\usepackage{bm}
\usepackage[colorlinks,citecolor=blue]{hyperref}
\usepackage{subcaption}
\usepackage{multirow}
\usepackage{threeparttable}
\usepackage{booktabs}
\usepackage{arydshln}
\usepackage{rotating}
\usepackage[numbers]{natbib}
\begin{document}
\title{Confounding Adjustment Methods for Multi-level Treatment Comparisons Under Lack of Positivity and Unknown Model Specification}

\author{S. Arona Diop \\ 
     serigne-arona.diop.1@ulaval.ca \\
     D\'epartement de math\'ematiques et de statistique \\
	Universit\'e Laval \\
   \And
  Thierry Duchesne\\
  thierry.duchesne@mat.ulaval.ca \\
  D\'epartement de math\'ematiques et de statistique \\
  Universit\'e Laval 
     \And
     Steven G Cumming \\
     stevec@sbf.ulaval.ca \\
	 D\'epartement des sciences du bois et de la for\^et \\
	 Universit\'e Laval 
			\And
     Awa Diop \\
     awa.diop.2@ulaval.ca \\
     D\'epartement de m\'edecine sociale et pr\'eventive \\
			Universit\'e Laval \\			
			\And
    Denis Talbot \\
    denis.talbot@fmed.ulaval.ca \\
    D\'epartement de m\'edecine sociale et pr\'eventive \\
	Universit\'e Laval \\
}

\maketitle
\begin{center}''The final, definitive version of this paper has been published in Journal of Applied Statistics by Taylor \& Francis.\\ DOI: https://doi.org/10.1080/02664763.2021.1911966\\
\textcopyright[S. Arona Diop, Thierry Duchesne, Steven G. Cumming, Awa Diop, Denis Talbot].''\end{center}

\begin{abstract}
Imbalances in covariates between treatment groups are frequent in observational studies and can lead to biased comparisons. Various adjustment methods can be employed to correct these biases in the context of multi-level treatments ($>$ 2). However, analytical challenges, such as positivity violations and incorrect model specification, may affect their ability to yield unbiased estimates. Adjustment methods that present the best potential to deal with those challenges were identified: the overlap weights, augmented overlap weights, bias-corrected matching and targeted maximum likelihood. A simple variance estimator for the overlap weight estimators that can naturally be combined with machine learning algorithms is proposed. In a simulation study, we investigated the empirical performance of these methods as well as those of simpler alternatives, standardization, inverse probability weighting and matching. Our proposed variance estimator performed well, even at a sample size of 500. Adjustment methods that included an outcome modeling component performed better than those that only modeled the treatment mechanism. Additionally, a machine learning implementation was observed to efficiently compensate for the unknown model specification for the former methods, but not the latter. Based on these results, the wildfire data were analyzed using the augmented overlap weight estimator. With respect to effectiveness of alternate fire-suppression interventions, the results were counter-intuitive, indeed the opposite of what would be expected on subject-matter grounds. This suggests the presence in the data of unmeasured confounding bias. 
\end{abstract}

\quad

\begin{center}
\textbf{Keywords} : multi-level treatment, machine learning, confounding adjustment, plasmode simulation
\end{center}

\clearpage
\section{Introduction}

Many empirical studies seek to evaluate the effect of a treatment or intervention. This evaluation can be attempted using randomized or observational experiments. In the former, pre-randomization characteristics are expected to be similar across treatment groups. As such, outcome differences can be causally attributed to the treatment. However, randomized trials are often difficult to realize because they may be unethical, impractical, or untimely  \citep{hernan2017causal}. Thus, relying on observational experiments is often necessary. Unfortunately, imbalances in covariates between treatment groups are frequent in observational studies and can lead to biased treatment comparisons. This major challenge is known as confounding, in which differences in outcomes between treatment groups are due, at least in part, to systematic differences in baseline covariates.

Various methods can be used to correct confounding bias. The performance of these methods have been compared in multiple simulation studies in the case of a binary treatment \citep[e.g.][]{lunceford2004stratification, austin2007comparison, austin2008performance, austin2010performance, huber2013performance, li2013weighting, kreif2016evaluating, mao2019propensity} and for performing pairwise comparisons between multi-level treatments \citep[e.g.][]{feng2012generalized, yang2016propensity, yoshida2017matching, li2018propensity, scotina2019matching}. In summary, stratification was susceptible to produce estimates with important residual bias \citep{lunceford2004stratification, austin2007comparison, austin2008performance, austin2010performance}. Standardization is generally the best performing method in ideal situations, in terms of bias and mean-squared error (MSE) \citep{lunceford2004stratification, austin2008performance, huber2013performance, li2013weighting, kreif2016evaluating}. However, when the model used for standardization is incorrectly specified, biased estimates are produced \citep{lunceford2004stratification, li2013weighting, kreif2016evaluating, feng2012generalized}. Weighting methods were susceptible to yield biased estimates with large variance if the model used for constructing the weights was misspecified \citep{huber2013performance, kreif2016evaluating, yang2016propensity} or when there was a lack of overlap in the covariates distribution between treatment groups, a problem also known as positivity violation \citep{mao2019propensity, li2018propensity, scotina2019matching}. Trimming, which removes from the analysis observations in areas where there is a lack of overlap, has been observed to improve the performance of multiple methods in presence of positivity violations \citep{huber2013performance, li2018propensity}. The augmented inverse probability weighting estimator has been observed to have mixed performances. This estimator performed competitively to standardization in ideal situations and retained similar performances under model misspecifications \citep{lunceford2004stratification}, but it performed poorly under positivity violations \citep{li2013weighting, huber2013performance}. The targeted maximum likelihood estimator (TMLE) was observed to have good performances under ideal situations and under model misspecifications, in addition to being robust to positivity violations \citep{kreif2016evaluating}. Similar results were observed concerning bias-corrected matching \citep{huber2013performance, li2013weighting, kreif2016evaluating}. The use of machine learning algorithms was observed to be beneficial for preventing the bias attributable to model misspecifications for TMLE, BCM and standardization, but not necessarily for weighting \citep{kreif2016evaluating}.

Weighting estimators that are robust to positivity violations have also been developed in recent years \citep{li2013weighting, yoshida2017matching, li2018balancing, li2018propensity}. In particular, \cite{li2018balancing} consider a general class of weighting estimators for binary treatments, that notably include as specific cases standard weighting and weighting with trimmed weights. They theoretically derived, under certain conditions, the lowest variance estimator in this class, which they called the overlap weights. \cite{li2018propensity} extended these overlap weights to multi-level treatments. The overlap weights were observed to outperform the usual weighting estimator, with or without trimming, as well as matching estimators, both in terms of bias and MSE \citep{li2018propensity, mao2019propensity}. 

Overall, a few methods thus stand out according to their performance in terms of bias, MSE, and their robustness to model misspecifications or positivity violations: bias-corrected matching, TMLE and overlap weights. While bias-corrected matching and TMLE have been compared in simulation studies with binary treatments, they have never been compared together in scenarios with multi-level treatment or against the overlap weights. Which of these methods yield the best estimator of pairwise average treatment effects, notably in scenarios with model misspecifications or positivity violations, is thus unknown. 

Hence, the goal of the current study was to investigate the performance of TMLE, bias-corrected matching and overlap weights for performing pairwise comparisons between multi-level treatments, notably under lack of positivity and unknown model specification. The motivation for this work was a comparison of the effect of five fire-suppression interventions on wildfire growth in Alberta, Canada. 

The remainder of this article is structured as follows. In the next section, the notation used throughout the article is first introduced. Then, the methods being compared are formally presented, as well as the causal assumptions on which they rely. We also propose some methodological developments concerning the overlap weight estimators, including a convenient asymptotic variance estimator. In Section~3, we present the simulation study we have conducted to compare bias-corrected matching, TMLE, overlap weights, an outcome regression augmented version of the overlap weights as well as standardization, inverse probability weighting and matching as benchmark comparators. We notably examine the potential of machine learning implementations of these methods to deal with the unknown model specification. Section $4$, presents the analysis of the wildfire data, which is informed by the results of our simulation study. We conclude with a discussion in Section $5$.

\section{Notation and Methods}

Let $Y$ be the outcome of interest, $T\in\mathcal{T} = \{1, 2, ..., k\}$ be the multi-level treatment, and $\bm{X} = \{X_1, ..., X_p\} \in \mathcal{X}$ be a set of covariates, where $\mathcal{T}$ and $\mathcal{X}$ are the support of variables $T$ and $\bm{X}$, respectively. We a assume that a sample of independent observations $\{\bm{X}_i, T_i, Y_i\}$, with $i = 1, ..., n$, is drawn from a given population. We also denote by $Y_i(t)$ the potential outcome of observation $i$ under treatment level $t$, that is, the outcome that would have been observed for observation $i$ had the treatment taken level $t$, potentially contrary to the fact. Using this notation, the population average pairwise treatment effect comparing treatment levels $t$ and $t'$ is $\tau_{t,t'} = \mathbb{E}[Y(t) - Y(t')]$.

The fundamental problem of causal inference is that it is only possible to observe one potential outcome for each observation, the one corresponding to the factual treatment level. Since the other potential outcomes are missing, the causal effect $\tau_{t,t'}$ cannot be estimated from the observed data without making some assumptions. Henceforth, we make the following usual assumptions \citep[see e.g.][Chapter 3]{hernan2017causal}: (1)~$Y(t) \coprod T | \bm{X}$ for all $t \in \mathcal{T}$, (2)~$0 < P(T = t|\bm{X}) < 1$ for all $t \in \mathcal{T}$ and $\bm{X} \in \mathcal{X}$ and (3)~$T_i = t \Rightarrow Y_i = Y_i(t)$. The first assumption is exchangeability and indicates that treatment allocation is independent of the potential outcomes conditional on measured covariates. Informally, this means there are no factors that would simultaneously explain $T$ and $Y$ beyond what may be explained by $\bm{X}$. Assumption (2) is the positivity assumption that entails that each observation had a strictly positive probability of being assigned to any treatment level. The third is the consistency assumption and indicates that the observed outcome for observation $i$ is equal to its potential outcome under its factual treatment level. 

\subsection{Standardization}

A first possible estimator of $\tau_{t,t'}$ consists of computing a standardized mean difference of the outcome between treatment groups $t$ and $t'$ \citep[see e.g.][Chapter~13]{hernan2017causal}:
\begin{align*}
\hat{\tau}^{stan}_{t,t'} = \frac{1}{n}\sum_{i=1}^n \left[\hat{\mathbb{E}}(Y|T = t, \bm{X}_i) - \hat{\mathbb{E}}(Y|T = t', \bm{X}_i) \right],
\end{align*}
\noindent where $\hat{\mathbb{E}}(Y|T = t, \bm{X}_i)$ and $\hat{\mathbb{E}}(Y|T = t', \bm{X}_i)$ are the estimated expectations of the outcome under treatment level $t$ and $t'$, respectively, and covariates $\bm{X}_i$. These estimated expectations can be obtained using a parametric regression model of $Y$ conditional on $T$ and $\bm{X}$, such as a linear regression if $Y$ is continuous or a logistic regression if $Y$ is binary. More sophisticated methods, including machine-learning algorithms, could also be employed. Although a variance estimator for $\hat{\tau}^{stan}_{t,t'}$ exists \citep{zou2009assessment}, it can be computationally intensive since it involves a double sum over the observations' index. Alternatively, inferences can be produced using the nonparametric bootstrap. 

\subsection{Inverse probability weighting}

Inverse probability weighting estimators of $\tau_{t,t'}$ are inspired by the method proposed by \cite{horvitz1952generalization}. \cite{Robins1997} introduced inverse probability weighting estimators of the parameters of longitudinal marginal structural models, which include as a special case the following estimator of $\tau_{t,t'}$:
\begin{align*}
\hat{\tau}^{IPW}_{t,t'} = \frac{1}{n}\sum_{i=1}^n \left[\frac{I(T_i = t)Y_i - I(T_i = t')Y_i}{\hat{P}(T = t_i|\bm{X}_i)}\right],
\end{align*}  
\noindent where $I(\cdot)$ denotes the usual indicator function and $\hat{P}(T = t_i|\bm{X}_i)$ is the estimated probability that the treatment takes level $t_i$ conditional on $\bm{X}_i$. $\hat{P}(T = t_i|\bm{X}_i)$ can, for example, be obtained from a multinomial logistic regression. A conservative asymptotic variance estimator for $\hat{\tau}^{IPW}_{t,t'}$ is conveniently obtained using a sandwich estimator, treating $\hat{P}(T = t_i|\bm{X}_i)$ as known \citep{robins2000marginalb, lunceford2004stratification}.  

\subsection{Matching and bias-corrected matching}

\cite{scotina2019matching} present a matching estimator as well as a bias-corrected matching estimator based on the work of \cite{abadie2006large} and \cite{abadie2011bias}. This matching estimator first imputes the missing potential outcomes of each observation by the observed outcomes of observations from the other treatment group. More precisely, the potential outcome of observation $i$ under treatment $t$, $Y_i(t)$, for $t \neq t_i$, is imputed with the average of the observed outcome of the $m$ observations from treatment group $t$ whose covariates' values are closest to $\bm{X}_i$ according to some distance metric. The potential outcome $Y_i(t_i)$ is deemed observed ($Y_i(t_i) = Y_i$). This matching procedure is performed with replacement, allowing each observation to be used multiple times for imputing potential outcomes of different observations. Comparisons between treatment groups are then performed directly on the (observed/imputed) potential outcomes. 

More formally, let $A$ be some definite positive matrix and $||x||_A = (x'Ax)^{1/2}$ define the distance metric. Let $M_i^t$ denote the set of the $m$ observations ``closest'' to unit $i$ in treatment group $t \neq t_i$, that is, such that $\sum_{j \in M_i^t} ||X_j - X_i||$ is minimal and $\sum_{j} I(j \in M_i^t) = m$. Then 
\begin{align*}
\hat{Y}_i(t) = \left\{
\begin{array}{cc}
Y_i, & \text{if } t = t_i, \\
\frac{1}{m} \sum_{j \in M_i^t} Y_j, & \text{if } t \neq t_i,
\end{array}
 \right.
\end{align*}
  
\noindent and the matching estimator is
\begin{align*}
\hat{\tau}_{t,t'}^{match} = \frac{1}{n} \sum_{i = 1}^n \left[\hat{Y}_i(t) - \hat{Y}_i(t')\right].
\end{align*}

\noindent Similar matching estimators for multi-level treatments had previously been introduced by \cite{frolich2004programme} and \cite{yang2016propensity}.  

The bias-corrected matching estimator proposed by \cite{scotina2019matching} uses a regression method for imputing the missing potential outcomes: 
\begin{align*}
\hat{Y}_i^{bc}(t) = \left\{
\begin{array}{cc}
Y_i, & \text{if } t = t_i, \\
\frac{1}{m} \sum_{j \in M_i^t} Y_j + \hat{\mathbb{E}}(Y|T = t, \bm{X}_i) - \hat{\mathbb{E}}(Y|T = t, \bm{X}_j), & \text{if } t \neq t_i.
\end{array}
 \right.
\end{align*}
\noindent The bias-corrected matching estimator is:
\begin{align*}
\hat{\tau}_{t,t'}^{BCM} = \frac{1}{n} \sum_{i = 1}^n \left[\hat{Y}_i^{bc}(t) - \hat{Y}_i^{bc}(t')\right].
\end{align*}   

The consistency of the matching estimator $\hat{\tau}_{t,t'}^{match}$ and the bias-corrected matching estimator $\hat{\tau}_{t,t'}^{BCM}$ relies on further conditions, in addition to the causal assumptions  presented earlier \citep[for more details, see][]{scotina2019matching}. Variance estimators of $\hat{\tau}_{t,t'}^{match}$ and of $\hat{\tau}_{t,t'}^{BCM}$ are provided in \cite{scotina2019matching}.

\subsection{Targeted maximum likelihood}

Targeted maximum likelihood estimation is a general methodology introduced by \cite{van2006targeted} for constructing doubly-robust semi-parametric efficient estimators. Double-robustness entails that the estimator is consistent for $\tau_{t,t'}$ if either the outcome model or the treatment model component of the algorithm described below is consistent, but not necessarily both. Moreover, the TMLE has minimal variance among the class of semiparametric estimators when both models are correctly specified. \cite{luque2018targeted} provide an introductory tutorial on TMLE.  

An algorithm for obtaining a TMLE, if $Y$ is binary, is as follows \citep{siddique2019causal}.
\begin{itemize}
	\item For $j \in \{t, t'\}$
		\begin{enumerate}
			\item Denote $Q^0(j, \bm{X}_i) = \hat{\mathbb{E}}(Y|T = j, \bm{X}_i)$
			\item Define weights $w_i(j) = \frac{I(T_i = j)}{\hat{P}(T = j|\bm{X}_i)}$
			\item Run a logistic regression of $Y$ with an intercept, $logit[Q^0(j,\bm{X})]$ as an offset term and weights $w_i(j)$. Denote the estimated intercept term as $\hat{\epsilon}$
			\item Let $Q^1(j, \bm{X}_i) = expit\{logit[Q^0(j, \bm{X}_i)] + \hat{\epsilon}\}$
		\end{enumerate}
\end{itemize}
\noindent The TMLE estimator is then
\begin{align*}
\hat{\tau}_{t,t'}^{TMLE} = \frac{1}{n} \sum_{i = 1}^n \left[Q^1(t, \bm{X}_i) - Q^1(t', \bm{X}_i)\right].
\end{align*} 

If $Y$ is a continuous variable, it is recommended to rescale $Y$ and $Q^0(j, \bm{X})$ such that values lie between 0 and 1 before performing step 3 \citep[][Chapter 7]{kreif2016evaluating, van2011targeted}. Let $a$ and $b$ be the known limits of $Y$, then the rescaled outcome and expectations are $Y^* = \frac{Y - a}{b - a}$ and $Q^{0,*}(j, \bm{X}) = \frac{Q^{0}(j, \bm{X}) - a}{b - a}$. The causal effect $\hat{\tau}_{t,t'}^{TMLE}$ is computed after back-transforming $Q^1(t, \bm{X}_i)$ and $Q^1(t', \bm{X}_i)$ into the original scale.  

An asymptotic estimator for the variance of $\hat{\tau}_{t,t'}^{TMLE}$ is based on the sample variance of the efficient influence function \citep[see e.g.][Chapter~5]{van2011targeted}. 

\subsection{Overlap weights \label{sec:OW}}

When using the overlap weights, each observation is attributed a weight $w_i = h(\bm{X}_i)/\hat{P}(T = t_i|\bm{X}_i)$, where $h(\bm{X}_i) = \left[\sum_{l = 1}^k 1/\hat{P}(T = l|\bm{X}_i)\right]^{-1}$. The overlap weight estimator is then
\begin{align*}
\hat{\tau}_{t,t'}^{OW} = \frac{\sum_{i = 1}^n I(T_i = t) Y_i w_i }{\sum_{i = 1}^n I(T_i = t) w_i} - \frac{\sum_{i = 1}^n I(T_i = t') Y_i w_i }{\sum_{i = 1}^n I(T_i = t') w_i}.
\end{align*} 
This specific choice of $h(\bm{X}_i)$ yields an estimator with minimal variance, assuming homoscedastic residual variances of the potential outcomes \citep{li2018propensity}. 

It is important to note that $\hat{\tau}_{t,t'}^{OW}$ is not generally a consistent estimator for the population average pairwise treatment effect $\tau_{t,t'}$. In fact, the estimand of $\hat{\tau}_{t,t'}^{OW}$ is the so-called average pairwise treatment effect in the overlap population:
\begin{align*}
\tau_{t,t'}^* = \frac{\int_{\bm{X}\in\mathcal{X}}\left\{\mathbb{E}[Y(t)|\bm{X}] - \mathbb{E}[Y(t')|\bm{X}]\right\} f(\bm{X})h(\bm{X})\mu(d\bm{X})}{\int_{\bm{X}\in\mathcal{X}}f(\bm{X})h(\bm{X})\mu(d\bm{X})},
\end{align*} 
\noindent where $f(\bm{X})$ is the marginal density of $\bm{X}$ with respect to an appropriate measure $\mu$. This corresponds to a weighted average treatment effect, where $h(\bm{X})$ is the weight function. 

A variance estimator for $\hat{\tau}_{t,t'}^{OW}$ that accounts for the estimation of the weights is provided by \cite{li2018propensity}. However, similar to the common approach for inverse probability weighting estimators, a simpler conservative variance estimator for $\hat{\tau}_{t,t'}^{OW}$ is obtained by considering the weights as known. This variance estimator corresponds to the empirical variance of the influence function of $\tau_{t,t'}^*$ \citep[given by][]{li2018propensity}, scaled by $1/n$: 
\begin{align*}
\widehat{Var}(\hat{\tau}_{t,t'}^{OW}) = & \frac{1}{n^2}\sum_{i=1}^n \left[\frac{I(T_i = t) (Y_i - \hat{\tau}_t) w_i}{\frac{1}{n}\sum_{i=1}^n h(\bm{X}_i)} - \frac{I(T_i = t')(Y_i - \hat{\tau}_{t'}) w_i}{\frac{1}{n}\sum_{i=1}^n h(\bm{X}_i)} \right]^2, 
\end{align*}
\noindent where $\hat{\tau}_t = \frac{\sum_{i = 1}^n I(T_i = t) Y_i w_i }{\sum_{i = 1}^n I(T_i = t) w_i}$ is the treatment $t$ weighted mean. This variance estimator is motivated by the fact the behavior of a semiparametric estimator is asymptotically the same as the one from its influence function \citep[][Chapter 5]{van2006targeted}. Effect estimates and inferences based on this variance estimator correspond to those produced by standard software when running a weighted generalized estimating equation regression with the treatment as the sole predictor and employing the robust variance estimator \citep[for example, using \texttt{proc genmod} in SAS with a \texttt{repeated} statement, or using \texttt{geeglm} in the \texttt{geepack} package in R,][]{halekoh2006r}. 

As indicated by \cite{li2018propensity} it is possible to construct an outcome-regression augmented version of the overlap weight estimator:
\begin{align*}
\hat{\tau}_{t,t'}^{A-OW} =& \hat{\tau}_{t,t'}^{OW} + \frac{\sum_{i=1}^n h(\bm{X}_i) \left[\hat{\mathbb{E}}(Y|T = t, \bm{X}_i) - \hat{\mathbb{E}}(Y|T = t', \bm{X}_i)\right]}{\sum_{i=1}^n h(\bm{X}_i)} \\
& - \frac{\sum_{i=1}^n I(T_i = t) \hat{\mathbb{E}}(Y|T = t, \bm{X}_i)w_i }{\sum_{i=1}^n I(T_i = t) w_i} + \frac{\sum_{i=1}^n I(T_i = t') \hat{\mathbb{E}}(Y|T = t', \bm{X}_i) w_i}{\sum_{i=1}^n I(T_i = t') w_i}.  
\end{align*}
\noindent This estimator is semiparametric efficient for estimating $\tau_{t,t'}^*$ when both the treatment model $\hat{P}(T = t|\bm{X})$ and the outcome model $\hat{\mathbb{E}}(Y|T, \bm{X}_i)$ are correctly specified \citep{li2018propensity}.

Following \cite{mao2019propensity}, we can show that $\hat{\tau}_{t,t'}^{A-OW}$ has a property \emph{analogue} to double-robustness.
\begin{theorem}
The estimator $\hat{\tau}_{t,t'}^{A-OW}$ is consistent for $\tau_{t,t'}^*$ when the treatment model is correctly specified, whether the outcome model is correctly specified or not. When the outcome model is correctly specified, but the treatment model is misspecified,  $\hat{\tau}_{t,t'}^{A-OW}$ is consistent for 
\begin{equation*}
\tilde \tau_{t,t'}^* = \frac{\int_{\bm{X}\in\mathcal{X}}\left\{\mathbb{E}[Y(t)|\bm{X}] - \mathbb{E}[Y(t')|\bm{X}]\right\} f(\bm{X})\tilde{h}(\bm{X})\mu(d\bm{X})}{\int_{\bm{X}\in\mathcal{X}}f(\bm{X})\tilde{h}(\bm{X})\mu(d\bm{X})},
\end{equation*}
where $\tilde{h}(\bm{X})$ is the estimand of $\left[\sum_{l = 1}^k 1/\hat{P}(T = l|\bm{X}_i)\right]^{-1}$ under the misspecified $\hat{P}(T = l|\bm{X}_i)$. 
\end{theorem}
\noindent The proof of Theorem~1 is provided in Web Appendix~A. 

Based on the efficient influence function derived by \cite{mao2019propensity}, an asymptotic variance estimator for $\hat{\tau}_{t,t'}^{A-OW}$ can conveniently be obtained by computing the empirical variance of the efficient influence function, scaled by a factor $1/n$ 
\begin{align*}
\widehat{Var}&(\hat{\tau}_{t,t'}^{A-OW}) = \frac{1}{n^2} \sum_{i=1}^n \left\{\frac{h(\bm{X}_i)}{\frac{1}{n}\sum_{i=1}^n h(\bm{X}_i)} \left[\hat{\mathbb{E}}(Y|T = t, \bm{X}_i) - \hat{\mathbb{E}}(Y|T=t', \bm{X}_i) \right. \right. \\ 
&+ \left. \left. \frac{I(T_i = t)\{Y_i - \hat{\mathbb{E}}(Y|T=t, \bm{X}_i)\}}{\hat{P}(T=t|\bm{X}_i)} - \frac{I(T_i = t')\{Y_i - \hat{\mathbb{E}}(Y|T=t', \bm{X}_i)\}}{\hat{P}(T=t'|\bm{X}_i)} \right] - \hat{\tau}_{t,t'}^{A-OW}  \right\}^2.
\end{align*}

\section{A simulation study}
\label{sec:sim}
We now describe the simulation study we have conducted to compare the empirical performance of the methods presented in the previous section.  

\subsection{Monte Carlo simulation}
\subsubsection{Simulation design and scenarios}
\label{metrics}
For our data-generating process, we considered three covariates $X_1$, $X_2$, $X_3$ arbitrarily generated as follow: $X_1 \sim \mathcal{N}(0,\,1)$, $X_2  \sim \mathcal{N}(2X_1,\,1)$ and $X_3  \sim \mathcal{B}ernoulli(0.4)$. The treatment variable $T$ was simulated according to a multinomial logistic regression with three levels ($1$, $2$ and $3$), where the first was the reference category. The probability of membership in each group was given by: 
\begin{align*}
\pi_1 = \left[1+ \exp(\beta_{21}X_1 + \beta_{22}X_2 + \beta_{23} X_3 + \beta_{24} X_1 X_3 + \beta_{25} X_1^2) \right. \\
 \left.+\exp(\beta_{31}X_1 +\beta_{32}X_2 + \beta_{33} X_3  + \beta_{34} X_1 X_3 + \beta_{35} X_1^2)\right]^{-1} \text{, } \\
\pi_2 =\exp(\beta_{21}X_1 +\beta_{22}X_2 + \beta_{23} X_3 + \beta_{24} X_1 X_3 + \beta_{25} X_1^2)\pi_1 \text{, } \\
\pi_3 =\exp(\beta_{31}X_1 +\beta_{32}X_2 + \beta_{33} X_3 + \beta_{34} X_1 X_3 + \beta_{35} X_1^2)\pi_1 \text{.}
\end{align*}
\noindent The outcome $Y$ was generated from a normal distribution $Y \sim \mathcal{N}(\mu_3,\,1)$, where $\mu_3 = \gamma_0 + \gamma_1X_1 + \gamma_2X_2 + \gamma_3X_3 + \gamma_4 X_2 X_3 + \gamma_5 X_2^2 + \lambda_1 I(T=2) + \lambda_2I(T=3)$.

We defined four simulation scenarios where the association between the covariates and the treatment was either weak ($T-$) or strong ($T+$), and the association between the covariates and the outcome was either weak ($Y-$) or strong ($Y+$). Parameters $\lambda_1$ and $\lambda_2$ represent the true treatment effects of level $2$ and $3$ as compared to level $1$ ($\tau_{21}$ and $\tau_{31}$), respectively, and were set to $\lambda_1 = 1.0$ and $\lambda_2 = 1.5$. These Monte Carlo scenarios were devised such that estimand of the overlap weights is the same as the one from the other methods. 

The specific choice of the values for $\beta$ and $\gamma$ was made to attain the following desiderata: 1) in Scenario $T-Y-$, the confounding bias for both $\lambda$ parameters is between 10\% and 30\%, 2) the bias for Scenarios $T-Y+$ and $T+Y-$ is between 40\% and 60\%, 3) the bias for Scenario $T+Y+$ is above 100\%, 4) there is a good overlap of the treatment probabilities distribution between treatment groups in Scenarios $T-$, and 5) there is a poor overlap in Scenarios $T+$.    

In Scenarios $T-$, the $\beta$ values were $\beta_{21} = -0.2$, $\beta_{22} = 0.2$, $\beta_{23} = 0.2$, $\beta_{24} = 0.1$, $\beta_{25} = 0.1$, $\beta_{31} = 0.2$, $\beta_{32} = 0.1$, $\beta_{33} = -0.2$, $\beta_{34} = 0.1$, $\beta_{35} = 0.1$. In Scenarios $T+$, they were $\beta_{21} = -0.8$, $\beta_{22} = 0.8$, $\beta_{23} = 0.8$, $\beta_{24} = 0.2$, $\beta_{25} = 0.2$, $\beta_{31} = 0.5$, $\beta_{32} = 0.5$, $\beta_{33} = -0.8$, $\beta_{34} = 0.2$, $\beta_{35} = 0.2$. In Scenarios $Y-$, the $\gamma$ values were $\gamma_0 = 0$, $\gamma_1 = 0.2$, $\gamma_2 = 0.2$, $\gamma_3 = 0.2$, $\gamma_4 = 0.1$ and $\gamma_5 = 0.1$. In Scenarios $Y+$, they were $\gamma_0 = 0$, $\gamma_1 = 0.5$, $\gamma_2 = 0.5$, $\gamma_3 = 0.5$, $\gamma_4 = 0.2$ and $\gamma_5 = 0.2$. 

For each scenario, we simulated $1000$ independent data sets of sample size $500, 1000$ and $2000$ under the above conditions. We then estimated $\tau_{21}$ and $\tau_{31}$ utilizing the standardization ($stan$), inverse probability weighting ($IPW$), matching ($match$), bias-corrected matching ($BCM$), targeted maximum likelihood ($TMLE$), overlap weights ($OW$) and augmented overlap weights ($A-OW$) estimators presented previously. First, the estimators were implemented using correctly specified parametric models, that is, a linear regression of $Y$ on $T$, $X_1$, $X_2$, $X_3$, $X_2 X_3$ and $X_2^2$, and a multinomial logistic regression of $T$ on $X_1$, $X_2$, $X_3$, $X_1 X_3$ and $X_1^2$. This allows investigating the performance of the estimators under ideal circumstances. Next, the estimators were implemented using parametric models that include only main terms, excluding interaction or quadratic terms. This corresponds to a common implementation since the correct model is unknown in practice. Finally, the estimators were implemented using machine learning approaches. For the outcome, we used a Super Learner. The Super Learner is an ensemble method that yields an estimate corresponding to a weighted average of the prediction of multiple procedures, where the weights are determined using cross-validation \citep{van2007super}. The prediction procedures we used were a main terms only linear regression, a main, interactions and quadratic terms linear regression, and a generalized additive model with cubic splines. This was performed using the \texttt{SuperLearner} package in R \citep{Polley2018SuperLearner}. Because this package does not accommodate multinomial dependent variables, we used a polychotomous regression and multiple classification with the R package \texttt{polspline} \citep{koop1997polychotomous, Kooperberg2019polspline} to model the exposure. This approach uses linear splines and their tensor products to produce predictions. 

We computed the following measures to assess and compare the performance of the estimators: bias, standard deviation (Std), square-root of the mean-squared error (RMSE) and the proportion of the replications in which 95\% confidence intervals included the true value of the treatment effect (Coverage CI). For $stan$, inferences were produced using a nonparametric bootstrap estimate of the variance with 200 samples. For $IPW$, $TMLE$, $OW$, and $A-OW$, the influence function variance estimator was used. For $match$ and $BCM$, we used the variance estimator proposed by \cite{scotina2019matching}. To assess bias levels, we calculated an unadjusted estimate of the treatment effect.
 
\subsubsection{Simulation results}

Tables \ref{tab:est_sim1}, \ref{tab:est_sim2}, \ref{tab:est_sim3} and \ref{tab:est_sim4} compare the aforementioned performance criteria across methods for $n = 1000$. Similar results were obtained for $n = 500$ and $n = 2000$. A detailed presentation of these results is available in Web Appendix~B. A plot depicting the overlap in the treatment probabilities distribution according to treatment group and scenarios is also available in Web Appendix~B.

Under a correct parametric specification, all methods managed to almost completely eliminate the bias in Scenarios $T-Y-$ and $T-Y+$. In Scenarios with positivity violations ($T+Y-$ and $T+Y+$), $IPW$ and $match$ yielded results with substantial residual bias, while the other methods still achieved close to unbiased estimation. In all scenarios, $stan$ produced the estimates with lowest standard deviation and RMSE, followed closely by $A-OW$, and by TMLE in Scenarios $T-Y-$ and $T-Y+$. In Scenarios $T+Y-$ and $T+Y+$, the standard deviation and RMSE of $IPW$ and $match$ were drastically larger than those of the other methods. The coverage of confidence intervals of all methods was close to 95\% or slightly conservative, except for $IPW$ and $match$ in Scenarios $T+Y-$ and $T+Y+$ where confidence intervals included the true effect much less often than expected. 

When using an incorrect parametric specification, important bias reduction was still achieved, but residual confounding bias remained for all adjustment methods. In Scenarios $T-Y-$ and $T-Y+$, $BCM$ and $match$ were the methods that reduced the bias the most and were the only methods that yielded confidence intervals with appropriate coverage. In Scenarios $T+Y-$ and $T+Y+$, $A-OW$ and $BCM$ were the methods that produced the most important bias reduction and the coverage of their confidence intervals were the closest, yet inferior, to 95\%.

When a machine learning implementation was used, $stan$, $BCM$, $TMLE$ and $A-OW$ produced results very similar to those obtained under the correct parametric implementations, with regards to bias, standard deviation, RMSE and coverage of confidence intervals. However, $IPW$, $match$ and $OW$ produced biased results with increased standard deviations and RMSE. The coverage of their confidence intervals was also generally below the expected level, except for $match$ that retained appropriate coverage in Scenarios $T-Y-$ and $T-Y+$.  

In all scenarios, we noticed that the bias of $match$ and $BCM$ tended to decrease with increased sample size, even under an incorrect parametric implementation. 

\begin{table}[hbpt]
	\centering
	\caption{Estimate of the treatment effect in the Scenario with a weak association between covariates and treatment, a weak association between covariates and outcome ($T-Y-$), and a sample size of 1000.} 
	\label{tab:est_sim1}
		\begin{tabular}{llcccccccc}
			\\[-1.8ex]\hline 
			\hline \\[-1.8ex] 
			\multirow{2}{*}{Implementation} & \multicolumn{1}{c}{\multirow{2}{*}{Approach}}  & \multicolumn{2}{c}{Bias} & \multicolumn{2}{c}{Std}& \multicolumn{2}{c}{RMSE}& \multicolumn{2}{c}{Coverage IC} \\
			\cmidrule(lr){3-4} \cmidrule(lr){5-6}  \cmidrule(lr){7-8}  \cmidrule(lr){9-10}  
			&  &  $\hat{\tau}_{21}$  &  $\hat{\tau}_{31}$ & $\hat{\tau}_{21}$  & $\hat{\tau}_{31}$ & $\hat{\tau}_{21}$ & $\hat{\tau}_{31}$ & $\hat{\tau}_{21}$ & $\hat{\tau}_{31}$ \\ 			
			\hline \\[-1.8ex] 
\multirow{8}{*}{Cor.param.} & \multicolumn{1}{c}{Crude} &  0.240 & 0.347 & 0.106 & 0.112 & 0.262 & 0.365 & 0.422 & 0.129 \\
                            & \multicolumn{1}{c}{stan}  &  0.001 & 0.002 & 0.081 & 0.082 & 0.081 & 0.082 & 0.948 & 0.960 \\ 
                            & \multicolumn{1}{c}{IPW}   &  0.003 & 0.006 & 0.102 & 0.103 & 0.102 & 0.103 & 0.983 & 0.985 \\ 
                            & \multicolumn{1}{c}{match} &  0.018 & 0.019 & 0.099 & 0.099 & 0.101 & 0.100 & 0.956 & 0.961 \\
                            & \multicolumn{1}{c}{BCM}   &  0.003 & 0.003 & 0.098 & 0.098 & 0.098 & 0.098 & 0.958 & 0.962 \\ 
                            & \multicolumn{1}{c}{TMLE}  &  0.002 & 0.003 & 0.084 & 0.085 & 0.084 & 0.085 & 0.953 & 0.960 \\ 
                            & \multicolumn{1}{c}{OW}    & -0.000 & 0.002 & 0.091 & 0.092 & 0.091 & 0.092 & 0.985 & 0.978 \\ 
                            & \multicolumn{1}{c}{A-OW}  &  0.001 & 0.002 & 0.083 & 0.084 & 0.083 & 0.084 & 0.940 & 0.943 \\ 
			\hdashline
\multirow{7}{*}{Inc.param.} & \multicolumn{1}{c}{stan}  & 0.046 & 0.057 & 0.096 & 0.097 & 0.107 & 0.112 & 0.922 & 0.905 \\  
                            & \multicolumn{1}{c}{IPW}   & 0.092 & 0.087 & 0.104 & 0.104 & 0.140 & 0.136 & 0.882 & 0.898 \\  
                            & \multicolumn{1}{c}{match} & 0.032 & 0.034 & 0.097 & 0.099 & 0.102 & 0.105 & 0.953 & 0.955 \\
                            & \multicolumn{1}{c}{BCM}   & 0.024 & 0.022 & 0.096 & 0.098 & 0.099 & 0.100 & 0.959 & 0.958 \\  
                            & \multicolumn{1}{c}{TMLE}  & 0.076 & 0.077 & 0.100 & 0.102 & 0.126 & 0.128 & 0.890 & 0.902 \\  
                            & \multicolumn{1}{c}{OW}    & 0.083 & 0.078 & 0.100 & 0.100 & 0.130 & 0.127 & 0.893 & 0.908 \\  
                            & \multicolumn{1}{c}{A-OW}  & 0.068 & 0.069 & 0.097 & 0.098 & 0.118 & 0.120 & 0.870 & 0.885 \\  
			\hdashline
\multirow{7}{*}{M.Learning} & \multicolumn{1}{c}{stan}         & 0.001 & 0.005 & 0.084 & 0.085 & 0.084 & 0.085 & 0.926 & 0.941 \\ 
                            & \multicolumn{1}{c}{IPW}          & 0.101 & 0.079 & 0.106 & 0.105 & 0.146 & 0.132 & 0.857 & 0.905 \\ 
                            & \multicolumn{1}{c}{match}    & 0.035 & 0.020 & 0.109 & 0.110 & 0.115 & 0.111 & 0.940 & 0.954 \\ 
                            & \multicolumn{1}{c}{BCM}      & -0.003 & 0.001 & 0.101 & 0.102 & 0.101 & 0.102 & 0.958 & 0.965 \\ 
                            & \multicolumn{1}{c}{TMLE}         & 0.004 & 0.007 & 0.085 & 0.086 & 0.085 & 0.086 & 0.957 & 0.955 \\ 
                            & \multicolumn{1}{c}{OW}           & 0.091 & 0.070 & 0.102 & 0.102 & 0.137 & 0.124 & 0.868 & 0.907 \\ 
                            & \multicolumn{1}{c}{A-OW}  & 0.002 & 0.006 & 0.084 & 0.085 & 0.085 & 0.086 & 0.933 & 0.941 \\ 
			\hline \\
		\end{tabular}%
\begin{small}

\begin{textit}\\
Legend: Cor.param=correct parametric models, Inc.param=incorrect parametric models, M.Learning=machine learning, Crude=Unadjusted, stan=standardization, IPW=inverse probability weighting, match=matching, BCM=bias-corrected matching, TMLE=targeted maximum likelihood, OW=overlap weights, A-OW=augmented overlap weights. 
\end{textit}
\end{small}
\end{table} 

\begin{table}[hbpt]
	\centering
	\caption{Estimate of the treatment effect in the Scenario with a strong association between covariates and treatment, a weak association between covariates and outcome ($T+Y-$), and a sample size of 1000.} 
	\label{tab:est_sim2}
		\begin{tabular}{llcccccccc}
			\\[-1.8ex]\hline 
			\hline \\[-1.8ex] 
			\multirow{2}{*}{Implementation} & \multicolumn{1}{c}{\multirow{2}{*}{Approach}}  & \multicolumn{2}{c}{Bias} & \multicolumn{2}{c}{Std}& \multicolumn{2}{c}{RMSE}& \multicolumn{2}{c}{Coverage IC} \\
			\cmidrule(lr){3-4} \cmidrule(lr){5-6}  \cmidrule(lr){7-8}  \cmidrule(lr){9-10}  
			&  &  $\hat{\tau}_{21}$  &  $\hat{\tau}_{31}$ & $\hat{\tau}_{21}$  & $\hat{\tau}_{31}$ & $\hat{\tau}_{21}$ & $\hat{\tau}_{31}$ & $\hat{\tau}_{21}$ & $\hat{\tau}_{31}$ \\ 			
			\hline \\[-1.8ex] 
\multirow{8}{*}{Cor.param.} & \multicolumn{1}{c}{Crude} & 0.467 & 0.790 & 0.094 & 0.117 & 0.477 & 0.799 & 0.001 & 0.000 \\ 
                            & \multicolumn{1}{c}{stan}  & 0.005 & 0.001 & 0.083 & 0.094 & 0.083 & 0.094 & 0.955 & 0.967 \\ 
                            & \multicolumn{1}{c}{IPW}   & 0.039 & 0.032 & 0.301 & 0.308 & 0.303 & 0.310 & 0.870 & 0.898 \\ 
                            & \multicolumn{1}{c}{match} & 0.102 & 0.092 & 0.121 & 0.140 & 0.158 & 0.168 & 0.862 & 0.889 \\ 
                            & \multicolumn{1}{c}{BCM}   & 0.022 & 0.004 & 0.120 & 0.137 & 0.122 & 0.137 & 0.958 & 0.960 \\ 
                            & \multicolumn{1}{c}{TMLE}  & 0.006 & 0.002 & 0.101 & 0.118 & 0.102 & 0.118 & 0.952 & 0.960 \\ 
                            & \multicolumn{1}{c}{OW}    & 0.009 & 0.004 & 0.104 & 0.116 & 0.105 & 0.116 & 0.967 & 0.975 \\ 
                            & \multicolumn{1}{c}{A-OW}  & 0.006 & 0.003 & 0.096 & 0.107 & 0.096 & 0.107 & 0.948 & 0.959 \\ 
			\hdashline
\multirow{7}{*}{Inc.param.} & \multicolumn{1}{c}{stan}  & -0.121 & -0.039 & 0.098 & 0.113 & 0.155 & 0.120 & 0.785 & 0.928 \\ 
                            & \multicolumn{1}{c}{IPW}   &  0.209 &  0.181 & 0.174 & 0.195 & 0.272 & 0.266 & 0.621 & 0.755 \\  
                            & \multicolumn{1}{c}{match} &  0.132 &  0.118 & 0.121 & 0.136 & 0.179 & 0.180 & 0.799 & 0.869 \\  
                            & \multicolumn{1}{c}{BCM}   &  0.091 &  0.045 & 0.119 & 0.136 & 0.149 & 0.143 & 0.887 & 0.940 \\  
                            & \multicolumn{1}{c}{TMLE}  &  0.159 &  0.112 & 0.131 & 0.155 & 0.206 & 0.191 & 0.767 & 0.892 \\  
                            & \multicolumn{1}{c}{OW}    &  0.117 &  0.099 & 0.109 & 0.118 & 0.160 & 0.154 & 0.812 & 0.865 \\  
                            & \multicolumn{1}{c}{A-OW}  &  0.065 &  0.062 & 0.101 & 0.116 & 0.120 & 0.131 & 0.898 & 0.920 \\
\hdashline
\multirow{7}{*}{M.Learning} & \multicolumn{1}{c}{stan}       & 0.005 & 0.014 & 0.090 & 0.105 & 0.090 & 0.106 & 0.938 & 0.940 \\ 
                            & \multicolumn{1}{c}{IPW}        & 0.159 & 0.138 & 0.185 & 0.200 & 0.244 & 0.244 & 0.720 & 0.801 \\ 
                            & \multicolumn{1}{c}{match}      & 0.124 & 0.111 & 0.122 & 0.138 & 0.174 & 0.177 & 0.819 & 0.875 \\ 
                            & \multicolumn{1}{c}{BCM}        & 0.010 & 0.008 & 0.118 & 0.135 & 0.119 & 0.136 & 0.953 & 0.956 \\ 
                            & \multicolumn{1}{c}{TMLE}       & 0.016 & 0.006 & 0.106 & 0.123 & 0.108 & 0.124 & 0.939 & 0.945 \\ 
                            & \multicolumn{1}{c}{OW}         & 0.080 & 0.067 & 0.110 & 0.121 & 0.136 & 0.138 & 0.893 & 0.912 \\ 
                            & \multicolumn{1}{c}{A-OW$^*$}   & 0.003 & -0.001 & 0.098 & 0.108 & 0.098 & 0.108 & 0.937 & 0.952 \\ 
			\hline \\
		\end{tabular}%
\begin{small}

\begin{textit}\\
Legend: Cor.param=correct parametric models, Inc.param=incorrect parametric models, M.Learning=machine learning, Crude=Unadjusted, stan=standardization, IPW=inverse probability weighting, match=matching, BCM=bias-corrected matching, TMLE=targeted maximum likelihood, OW=overlap weights, A-OW=augmented overlap weights. **: in 1 replication, the confidence intervals could not be computed. 
\end{textit}
\end{small}
\end{table}

\begin{table}[hbpt]
	\centering
	\caption{Estimate of the treatment effect in the Scenario with a weak association between covariates and treatment, a strong association between covariates and outcome ($T-Y+$), and a sample size of 1000.} 
	\label{tab:est_sim3}
		\begin{tabular}{llcccccccc}
			\\[-1.8ex]\hline 
			\hline \\[-1.8ex] 
			\multirow{2}{*}{Implementation} & \multicolumn{1}{c}{\multirow{2}{*}{Approach}}  & \multicolumn{2}{c}{Bias} & \multicolumn{2}{c}{Std}& \multicolumn{2}{c}{RMSE}& \multicolumn{2}{c}{Coverage IC} \\
			\cmidrule(lr){3-4} \cmidrule(lr){5-6}  \cmidrule(lr){7-8}  \cmidrule(lr){9-10}  
			&  &  $\hat{\tau}_{21}$  &  $\hat{\tau}_{31}$ & $\hat{\tau}_{21}$  & $\hat{\tau}_{31}$ & $\hat{\tau}_{21}$ & $\hat{\tau}_{31}$ & $\hat{\tau}_{21}$ & $\hat{\tau}_{31}$ \\ 			
			\hline \\[-1.8ex] 
\multirow{8}{*}{Cor.param.} & \multicolumn{1}{c}{Crude} &  0.564 & 0.819 & 0.173 & 0.192 & 0.590 & 0.841 & 0.125 & 0.009 \\
                            & \multicolumn{1}{c}{stan}  &  0.001 & 0.002 & 0.081 & 0.082 & 0.081 & 0.082 & 0.984 & 0.986 \\
                            & \multicolumn{1}{c}{IPW}   &  0.005 & 0.008 & 0.149 & 0.151 & 0.149 & 0.151 & 0.997 & 0.995 \\ 
                            & \multicolumn{1}{c}{match} &  0.037 & 0.039 & 0.110 & 0.110 & 0.116 & 0.117 & 0.983 & 0.979 \\
                            & \multicolumn{1}{c}{BCM}   &  0.004 & 0.002 & 0.106 & 0.107 & 0.106 & 0.107 & 0.985 & 0.990 \\ 
                            & \multicolumn{1}{c}{TMLE}  &  0.002 & 0.003 & 0.084 & 0.085 & 0.084 & 0.085 & 0.953 & 0.960 \\
                            & \multicolumn{1}{c}{OW}    & -0.001 & 0.002 & 0.111 & 0.115 & 0.111 & 0.115 & 0.999 & 0.998 \\ 
                            & \multicolumn{1}{c}{A-OW}  &  0.001 & 0.002 & 0.083 & 0.084 & 0.083 & 0.084 & 0.940 & 0.943 \\
			\hdashline
\multirow{7}{*}{Inc.param.} & \multicolumn{1}{c}{stan}  & 0.046 & 0.057 & 0.096 & 0.097 & 0.107 & 0.112 & 0.922 & 0.905 \\ 
                            & \multicolumn{1}{c}{IPW}   & 0.092 & 0.087 & 0.104 & 0.104 & 0.140 & 0.136 & 0.882 & 0.898 \\ 
                            & \multicolumn{1}{c}{match} & 0.032 & 0.034 & 0.097 & 0.099 & 0.102 & 0.105 & 0.953 & 0.955 \\ 
                            & \multicolumn{1}{c}{BCM}   & 0.024 & 0.022 & 0.096 & 0.098 & 0.099 & 0.100 & 0.959 & 0.958 \\ 
                            & \multicolumn{1}{c}{TMLE}  & 0.076 & 0.077 & 0.100 & 0.102 & 0.126 & 0.128 & 0.890 & 0.902 \\  
                            & \multicolumn{1}{c}{OW}    & 0.083 & 0.078 & 0.100 & 0.100 & 0.130 & 0.127 & 0.893 & 0.908 \\ 
                            & \multicolumn{1}{c}{A-OW}  & 0.068 & 0.069 & 0.097 & 0.098 & 0.118 & 0.120 & 0.870 & 0.885 \\ 
\hdashline
\multirow{7}{*}{M.Learning} & \multicolumn{1}{c}{stan} & 0.001 & 0.005 & 0.084 & 0.085 & 0.084 & 0.085 & 0.926 & 0.941 \\ 
                            & \multicolumn{1}{c}{IPW} & 0.101 & 0.079 & 0.106 & 0.105 & 0.146 & 0.132 & 0.857 & 0.905 \\ 
                            & \multicolumn{1}{c}{match} & 0.034 & 0.019 & 0.108 & 0.108 & 0.113 & 0.110 & 0.940 & 0.954 \\
                            & \multicolumn{1}{c}{BCM} & -0.003 & 0.001 & 0.099 & 0.101 & 0.099 & 0.101 & 0.960 & 0.965 \\ 
                            & \multicolumn{1}{c}{TMLE} & 0.004 & 0.007 & 0.085 & 0.086 & 0.085 & 0.086 & 0.957 & 0.955 \\ 
                            & \multicolumn{1}{c}{OW} & 0.091 & 0.070 & 0.102 & 0.102 & 0.137 & 0.124 & 0.868 & 0.907 \\ 
                            & \multicolumn{1}{c}{A-OW} & 0.002 & 0.006 & 0.084 & 0.085 & 0.085 & 0.086 & 0.933 & 0.941 \\ 
			\hline \\
		\end{tabular}%
\begin{small}
\begin{textit}\\
Legend: Cor.param=correct parametric models, Inc.param=incorrect parametric models, M.Learning=machine learning, Crude=Unadjusted, stan=standardization, IPW=inverse probability weighting, match=matching, BCM=bias-corrected matching, TMLE=targeted maximum likelihood, OW=overlap weights, A-OW=augmented overlap weights.
\end{textit}
\end{small}
\end{table}

\begin{table}[hbpt]
	\centering
	\caption{Estimate of the treatment effect in the Scenario with a strong association between covariates and treatment, a strong association between covariates and outcome ($T+Y+$), and a sample size of 1000.} 
	\label{tab:est_sim4}
		\begin{tabular}{llcccccccc}
			\\[-1.8ex]\hline 
			\hline \\[-1.8ex] 
			\multirow{2}{*}{Implementation} & \multicolumn{1}{c}{\multirow{2}{*}{Approach}}  & \multicolumn{2}{c}{Bias} & \multicolumn{2}{c}{Std}& \multicolumn{2}{c}{RMSE}& \multicolumn{2}{c}{Coverage IC} \\
			\cmidrule(lr){3-4} \cmidrule(lr){5-6}  \cmidrule(lr){7-8}  \cmidrule(lr){9-10}  
			&  &  $\hat{\tau}_{21}$  &  $\hat{\tau}_{31}$ & $\hat{\tau}_{21}$  & $\hat{\tau}_{31}$ & $\hat{\tau}_{21}$ & $\hat{\tau}_{31}$ & $\hat{\tau}_{21}$ & $\hat{\tau}_{31}$ \\ 			
			\hline \\[-1.8ex] 
\multirow{8}{*}{Cor.param.} & \multicolumn{1}{c}{Crude} & 1.180 & 1.937 & 0.146 & 0.198 & 1.189 & 1.947 & 0.000 & 0.000 \\ 
                            & \multicolumn{1}{c}{stan}  & 0.005 & 0.001 & 0.083 & 0.094 & 0.083 & 0.094 & 0.987 & 0.988 \\ 
                            & \multicolumn{1}{c}{IPW}   & 0.079 & 0.075 & 0.612 & 0.614 & 0.617 & 0.618 & 0.841 & 0.866 \\ 
                            & \multicolumn{1}{c}{match} & 0.213 & 0.214 & 0.143 & 0.161 & 0.256 & 0.267 & 0.747 & 0.796 \\ 
                            & \multicolumn{1}{c}{BCM}   & 0.039 & 0.007 & 0.131 & 0.147 & 0.136 & 0.147 & 0.976 & 0.983 \\ 
                            & \multicolumn{1}{c}{TMLE}  & 0.006 & 0.002 & 0.102 & 0.118 & 0.102 & 0.118 & 0.951 & 0.959 \\ 
                            & \multicolumn{1}{c}{OW}    & 0.012 & 0.006 & 0.131 & 0.142 & 0.131 & 0.143 & 0.990 & 0.990 \\ 
                            & \multicolumn{1}{c}{A-OW}  & 0.006 & 0.003 & 0.096 & 0.107 & 0.096 & 0.107 & 0.948 & 0.959 \\ 
			\hdashline
\multirow{7}{*}{Inc.param.} & \multicolumn{1}{c}{stan}  & -0.121 & -0.039 & 0.098 & 0.113 & 0.155 & 0.120 & 0.785 & 0.928 \\  
                            & \multicolumn{1}{c}{IPW}   &  0.209 &  0.181 & 0.174 & 0.195 & 0.272 & 0.266 & 0.621 & 0.755 \\  
                            & \multicolumn{1}{c}{match} &  0.132 &  0.118 & 0.121 & 0.136 & 0.179 & 0.180 & 0.799 & 0.869 \\  
                            & \multicolumn{1}{c}{BCM}   &  0.091 &  0.045 & 0.119 & 0.136 & 0.149 & 0.143 & 0.887 & 0.940 \\  
                            & \multicolumn{1}{c}{TMLE}  &  0.159 &  0.112 & 0.131 & 0.155 & 0.206 & 0.191 & 0.767 & 0.892 \\  
                            & \multicolumn{1}{c}{OW}    &  0.117 &  0.099 & 0.109 & 0.118 & 0.160 & 0.154 & 0.812 & 0.865 \\  
                            & \multicolumn{1}{c}{A-OW}  &  0.065 &  0.062 & 0.101 & 0.116 & 0.120 & 0.131 & 0.898 & 0.920 \\  

\hdashline
\multirow{7}{*}{M.Learning} & \multicolumn{1}{c}{stan} & 0.005 & 0.014 & 0.090 & 0.105 & 0.090 & 0.106 & 0.938 & 0.940 \\ 
                            & \multicolumn{1}{c}{IPW} & 0.159 & 0.138 & 0.185 & 0.200 & 0.244 & 0.244 & 0.720 & 0.801 \\ 
                            & \multicolumn{1}{c}{match} & 0.124 & 0.111 & 0.122 & 0.138 & 0.174 & 0.177 & 0.819 & 0.875 \\ 
                            & \multicolumn{1}{c}{BCM} & 0.010 & 0.008 & 0.118 & 0.135 & 0.119 & 0.136 & 0.953 & 0.956 \\ 
                            & \multicolumn{1}{c}{TMLE} & 0.016 & 0.006 & 0.106 & 0.123 & 0.108 & 0.124 & 0.939 & 0.945 \\ 
                            & \multicolumn{1}{c}{OW} & 0.080 & 0.067 & 0.110 & 0.121 & 0.136 & 0.138 & 0.893 & 0.912 \\ 
                            & \multicolumn{1}{c}{A-OW$^*$} & 0.003 & -0.001 & 0.098 & 0.108 & 0.098 & 0.108 & 0.937 & 0.952 \\ 
\hline \\
		\end{tabular}
\begin{small}
\begin{textit}\\
Legend: Cor.param=correct parametric models, Inc.param=incorrect parametric models, M.Learning=machine learning, Crude=Unadjusted, stan=standardization, IPW=inverse probability weighting, match=matching, BCM=bias-corrected matching, TMLE=targeted maximum likelihood, OW=overlap weights, A-OW=augmented overlap weights. *: in 1 replication, the confidence intervals could not be computed. 
\end{textit}
\end{small}
\end{table}

\subsection{Plasmode simulation}

\subsubsection{Plasmode datasets}
We used the data on forest fires in Alberta, Canada, as a basis for our plasmode simulation. These data are produced and published online by the Government of Alberta \citep[Wildfire Management Branch -- Alberta Agriculture and Forestry,][]{historicalwildfiredatabase}. The purpose of the analysis was to compare various interventions for fighting wildfires in Alberta on their probability of preventing the fire to grow after its initial assessment. We considered only fires caused by lightning from $1996$ to $2014$. Observations for which size at ``being held'' was smaller than the size at initial attack were also removed. ``Being held'' is defined as in \cite{tremblay2018survival} as a state when no further increase in size is expected.

Potential confounders considered in the plasmode simulation are the ecological region in which the fire occurred (Clear Hills Upland, Mid-Boreal Uplands, Other), the number of fires active at the time of initial assessment of each fire, fuel type (Boreal Spruce, Boreal Mixedwood - Green, other), month of the year the fire was first assessed (``May or June'', July, ``August, September or October''), how the fire was discovered (air patrol, lookout, unplanned), response time in hours between the moment the fire was reported to first assessment by forestry personnel, and the natural logarithm ($\ln$) of the size of the fire at the time of the initial attack. Response time was truncated at the 95th percentile of the distribution to limit the influence of extreme observations, which caused convergence problems of models in some replications of the simulation. Additional potential confounders were available but were not considered in the plasmode simulation. As will be explained shortly, the plasmode simulation was run on a reduced sample of the total data available. As such, including all potential confounders in the simulations caused convergence issues. 

Some levels of categorical variables having few instances were dropped, and observations in those categories were removed. The final database contained $8591$ observations, the seven covariates presented above, and the treatment variable indicating the method of intervention used by the firefighters to suppress the wildfire and the outcome variable. The categories for the treatment were heli-attack crew with helicopter but no rappel capability (HAC1H; 53.8\%), heli-attack crew with helicopter and rappel capability (HAC1R; 15.8\%), fire-attack crew with or without a helicopter and no rappel capability (HAC1F; 6.3\%), Air tanker (15.3\%) and Ground-based action (8.8\%). The binary response variable was 1 or 0 depending on whether the fire did or did not increase in size between initial attack and ``being held'' (n=1982 and 6645, respectively). 

For generating plasmode data, we first fitted a random forest regression of the outcome variable conditional on treatment and all seven covariates using the package \texttt{randomForest} in R with the default settings \citep{Liaw2002Classification}. Similarly, a random forest regression of the treatment according to the covariates was fitted. These models were then used to generate simulated treatment and outcome variables. The true causal effects were estimated using a Monte Carlo simulation. Briefly, for each observation in the complete population, we generated counterfactual outcomes that would have been observed under each possible treatment according to the true outcome model. The difference in risk of fire progression were then computed for each treatment level, as compared to HAC1H, which was used as the reference level. For the overlap weights, risk differences weighted according to $h(\bm{X})$ as defined in Section \ref{sec:OW} were computed. The true effects comparing Air tanker, Ground-based action, HAC1F and HAC1R to HAC1H in the entire population were, respectively, 0.099, -0.011, 0.012 and -0.003. In the overlap population, the true risk differences were 0.101, -0.024, 0.003 and -0.002. To simplify the presentation, both sets of parameters will be denoted as $\tau_{21}$, $\tau_{31}$, $\tau_{41}$ and $\tau_{51}$ in the results below. 

A plasmode simulated dataset was obtained as follows. First, we drew with replacement a random sample of size $2000$ from the original data. Then, for each observation, a new simulated treatment and a new simulated outcome were generated, with $P(T = t|\bm{X})$ and $P(Y = 1|T, \bm{X})$ determined by the fitted random forest models. Based on this process, we simulated $1000$ independent data sets and compared adjustments methods using the same performance metrics as in the Monte Carlo simulation (see \hyperref[metrics]{Section \ref{metrics}}). For each method, we considered either a parametric models implementation that included only main terms, or a same machine learning implementation. For the treatment, these implementations were the same as described in Section \ref{metrics}. The implementations for the outcome were also similar to those described in Section \ref{metrics}, but replacing linear regressions by logistic regressions. Since the outcome and treatment data were generated nonparametrically, the correct models are unknown.

\subsubsection{Simulation results}
A figure displaying the distribution of the treatment probabilities according to treatment groups is available in Web Appendix~B. While a good level of overlap between treatment groups was present, multiple observations had treatment probabilities close to 0, thus suggesting possible practical positivity violations. 

The results presented in Table \ref{tab:est_plasm_2000} show that a large amount of bias is present when no adjustment is performed, especially for $\tau_{21}$. When using a parametric implementation, most methods achieved a bias reduction for estimating all parameters, except $stan$ and $match$ that increased the bias for one parameter. Additionally, a substantial bias remained for estimating some parameters for $stan$, $match$, and $A-OW$. The RMSE of $stan$, $TMLE$, and $OW$ were smaller than those of the crude estimates for all parameters, whereas $IPW$, $match$, $BCM$, and $A-OW$ had an increased RMSE for at least one parameter. The lowest RMSE for all parameters was produced by $stan$. Most methods yielded 95\% confidence intervals with coverage rate close to the expected level for $\tau_{21}$, $\tau_{41}$ and $\tau_{51}$, but not for $\tau_{31}$. Only $IPW$ and $OW$ had close to adequate coverage for $\tau_{31}$. $match$ and $BCM$ yielded inadequate confidence intervals for all parameters, because of an underestimation of the true variance (results not shown). When using a machine learning implementation, the performance of most adjustment methods was overall marginally improved. Indeed, the bias and RMSE were generally smaller, and coverage closer to 95\%. For $IPW$, however, the results were somewhat worse. 

\begin{table}[hbpt]
	\centering
	\begin{sideways}
		\begin{minipage}{\textheight}
			\caption{Estimate of treatment effect in plasmode simulation using 2000 observations. True effects are ${\tau}_{21}=-0.099$,  ${\tau}_{31}=0.011$,  ${\tau}_{41}=-0.012$ and  ${\tau}_{51}=0.003$ for all methods except $OW$ and $A-OW$ and ${\tau}_{21}=-0.101$,  ${\tau}_{31}=0.024$,  ${\tau}_{41}=-0.003$ and  ${\tau}_{51}=0.002$ for the these two methods. HAC1H is the reference category . ${\tau}_{21}$, ${\tau}_{31}$, ${\tau}_{41}$ and ${\tau}_{51}$ refer to treatment effect associated to Air tanker, Ground-based action, HAC1F and  HAC1R, respectively.} 
			\label{tab:est_plasm_2000}
			\resizebox{\textwidth}{!}{%
				\begin{tabular}{llcccccccccccc}
					\\[-1.8ex]\hline 
					\hline \\[-1.8ex] 
					\multicolumn{1}{c}{\multirow{4}{*}{Implementation}} & \multicolumn{1}{c}{\multirow{4}{*}{Approach}} & \multicolumn{4}{c}{Bias}  &   \multicolumn{4}{c}{RMSE}& \multicolumn{4}{c}{Coverage IC} \\
					\cmidrule(lr){3-6} \cmidrule(lr){7-10}  \cmidrule(lr){11-14}   &  & $\widehat{\tau}_{21}$  & $\widehat{\tau}_{31}$
					&$\widehat{\tau}_{41}$  & $\widehat{\tau}_{51} $  &  $\widehat{\tau}_{21}$  & $\widehat{\tau}_{31}$
					&$\widehat{\tau}_{41}$  & $\widehat{\tau}_{51} $  &  $\widehat{\tau}_{21}$  & $\widehat{\tau}_{31}$
					&$\widehat{\tau}_{41}$  & $\widehat{\tau}_{51} $ \\
					\hline
\multirow{8}{*}{Parametric} & \multicolumn{1}{c}{Crude} & -0.164 & -0.027 & -0.021 & -0.035 & 0.168 & 0.045 & 0.050 & 0.045 & 0.002 & 0.897 & 0.941 & 0.786 \\ 
                            & \multicolumn{1}{c}{stan}  & 0.004  & 0.028  & 0.009  & -0.005 & 0.029 & 0.043 & 0.041 & 0.026 & 0.921 & 0.800 & 0.939 & 0.942 \\ 
                            & \multicolumn{1}{c}{IPW}   & -0.005 & 0.011  & 0.003  & -0.002 & 0.040 & 0.048 & 0.050 & 0.029 & 0.957 & 0.902 & 0.947 & 0.963 \\ 
                            & \multicolumn{1}{c}{match} & 0.040  & -0.004 & 0.023  & 0.004 & 0.050 & 0.043 & 0.052 & 0.042 & 0.279 & 0.428 & 0.335 & 0.398 \\ 
                            & \multicolumn{1}{c}{BCM}   & 0.004  & 0.003  & 0.002  & -0.013 & 0.045 & 0.052 & 0.053 & 0.055 & 0.406 & 0.353 & 0.324 & 0.326 \\ 
                            & \multicolumn{1}{c}{TMLE}  & -0.006 & 0.012  & 0.001  & -0.005 & 0.037 & 0.044 & 0.044 & 0.028 & 0.935 & 0.859 & 0.926 & 0.951 \\ 
                            & \multicolumn{1}{c}{OW}    & 0.001  & 0.005  & 0.009  & 0.004 & 0.041 & 0.045 & 0.049 & 0.033 & 0.955 & 0.902 & 0.936 & 0.966 \\ 
                            & \multicolumn{1}{c}{A-OW}  & -0.006 & 0.021  & 0.007  & -0.005 & 0.040 & 0.048 & 0.046 & 0.032 & 0.942 & 0.859 & 0.943 & 0.942 \\ 
\hdashline
\multirow{7}{*}{Machine Learning} & \multicolumn{1}{c}{stan} & 0.007   & 0.027   & 0.008   & -0.006 & 0.030 & 0.043 & 0.041 & 0.026 & 0.923 & 0.847 & 0.941 & 0.944 \\ 
                            & \multicolumn{1}{c}{IPW}        & -0.027  & 0.009   & -0.001  & -0.012 & 0.048 & 0.040 & 0.046 & 0.031 & 0.908 & 0.924 & 0.956 & 0.954 \\ 
                            & \multicolumn{1}{c}{match}      & 0.028   & -0.008  & 0.019   & 0.000 & 0.041 & 0.037 & 0.048 & 0.039 & 0.401 & 0.485 & 0.360 & 0.459 \\ 
                            & \multicolumn{1}{c}{BCM}        & 0.001   & 0.007   & 0.006   & -0.007 & 0.042 & 0.044 & 0.048 & 0.050 & 0.436 & 0.393 & 0.349 & 0.367 \\ 
                            & \multicolumn{1}{c}{TMLE}       & -0.003  & 0.018   & 0.007   & -0.004 & 0.035 & 0.043 & 0.042 & 0.027 & 0.932 & 0.837 & 0.925 & 0.942 \\ 
                            & \multicolumn{1}{c}{OW}         & -0.028  & 0.000   & -0.007  & -0.011 & 0.047 & 0.040 & 0.048 & 0.032 & 0.899 & 0.916 & 0.952 & 0.959 \\ 
                            & \multicolumn{1}{c}{A-OW}       & -0.005  & 0.025   & 0.009   & -0.004 & 0.035 & 0.047 & 0.045 & 0.028 & 0.935 & 0.824 & 0.930 & 0.948 \\ 
\hline \\[-1.8ex] 
				\end{tabular}%
			}
			\begin{small}
			
 \vspace{0.1cm}
\begin{textit}
Legend: Crude=Unadjusted, stan=standardization, IPW=inverse probability weighting, match=matching, BCM=bias-corrected matching, TMLE=targeted maximum likelihood, OW=overlap weights, A-OW=augmented overlap weights
\end{textit}
\end{small}
		\end{minipage}
	\end{sideways}
\end{table}

\section{Comparison of interventions for fighting wildfires in Alberta, Canada \label{SecDataAnalysis}}

\subsection{Context}
Wildfires have great economic, environmental and societal impacts. A key tool of fire management is fire suppression by initial attack teams who seek to limit the growth, and hence the final size, of the fire \citep{cumming2005effective}. These interventions by initial attack differ in terms of the size and training of the fire-fighting crews, and the mechanical resources provided to them. However, few studies have attempted estimating the causal effect of initial attack on wildfire growth. 

Recently, \citet{tremblay2018survival} compared initial attack interventions using public data on wildfire in Alberta, Canada. More aggressive interventions were associated with greater fire growth, opposite to the expected direction of the causal effect. These results suggest that important confounding by indication may be present in fire management agency records.  

\subsection{Data}
The data we considered are similar to those used for performing the plasmode simulation and those considered by \citet{tremblay2018survival}. More precisely, we considered all fires on provincial land or in other public lands caused by lighting between $2003$ and $2014$ recorded in the historical wildfire database of Alberta's Agriculture and Forestry ministry \citep{historicalwildfiredatabase}. 

The final data consisted of $5439$ observations. The fire-attack interventions being compared are the same as those in the plasmode simulation: heli-attack crew with helicopter but no rappel capability (HAC1H), heli-attack crew with helicopter and rappel capability (HAC1R), fire-attack crew with or without a helicopter and no rappel capability (HAC1F), Air tanker, and Ground-based action. The outcome of interest was whether the fire grew in size between initial assessment and ``being held.'' 

The following 11 potential confounders were considered: 1) Initial Spread Index, 2) Fire Weather Index, 3) year that the fire occurred, 4) how the fire was discovered (air patrol, lookout, unplanned), 5) ecological region in which the fire occurred (Clear Hills Upland, Mid-Boreal Uplands, Other), 6) fuel type at initial assessment (Boreal Spruce, Boreal Mixedwood - Green, Other), 7) period of day (AM or PM), 8) month of the year the fire was first assessed (``May or June'', July, ``August, September or October''), 9) response time in hours between the moment the fire was reported to first assessment by fire fighters, 10) the number of fires active at the time of initial assessment of each fire, and 11) the natural logarithm ($\ln$) of the size of the fire at the initial attack. The Initial Spread Index and the Fire Weather Index are used to predict the expected rate of progression of fires and fire danger, respectively, by Alberta's Wildfire Management Branch. These indexes are notably based on daily weather variables. To account for the fact that initial attack decisions would be based not only on current and recent weather, but also on future forecast, we used the values of these variables for the day of fire assessment, the two days prior to assessment and the two days following assessment (for a total of five variables for each index). More information on these variables is available in \cite{tremblay2018survival} and references therein. 

\subsection{Analysis}
Not all interventions would be reasonable choices to suppress some fires given their characteristics at the moment of assessment. As such, it seems more relevant to target the average effect of interventions only for those fires for which multiple options are reasonable. Based on this and on the results from our simulation study, we decided to use the augmented overlap-weights estimator with a machine learning implementation. The treatment probabilities were estimated using a polychotomous regression and the outcome expectations were estimated with the Super Learner using a main term generalized linear model, a generalized additive model with spline terms, and random forests as prediction procedures. The generalized linear model with interaction and quadratic terms was not considered as a prediction procedure because of the small sample size in some levels of categorical variables. All potential confounders were included as independent variables in both models. Year was considered as a categorical variable (12 levels). Inferences were produced using the variance estimator we introduced in Section~\ref{sec:OW}. P-values were adjusted for multiple comparisons with Holm's method \citep{holm1979simple}. 

\subsection{Results}
Characteristics of the considered fires according to the initial intervention that was chosen for suppressing the fire are presented in Web Appendix~C. Important imbalances between treatment groups were observed for calendar year, how the fire was discovered, ecological region, fuel type, response time, number of active fires and initial size of the fire, and some imbalance was observed for the fire weather index on the day of initial assessment. 

Adjusted associations between initial interventions for suppressing fires and the probability of fire growth are reported in Table~\ref{tabRes1}. Air tanker, which is the most aggressive intervention, is associated with the largest probability of fire growth. All interventions are associated with probabilities of fire growth similar to or greater than that of ground-based action, which we consider to be the least aggressive intervention. The various heli-attack crew interventions are all associated with similar probabilities of fire growth.

\begin{table}[!htbp] 
	\caption{Association between initial intervention used to suppress the fire and probability of fire growth between initial assessment and being held} 
	\label{tabRes1} 
	\centering 
\resizebox{\textwidth}{!}{%
\begin{tabular}{@{\extracolsep{5pt}} lccc} 
\\[-1.8ex]\hline 
\hline \\[-1.8ex] 
\multicolumn{1}{c}{Contrast} & \multicolumn{1}{c}{Risk difference} & \multicolumn{1}{c}{95\% confidence interval} & \multicolumn{1}{c}{Adjusted p-value} \\ 
\hline \\[-1.8ex] 
\multicolumn{1}{c}{Air tanker vs HAC1H}               & 0.107 & (0.083, 0.131) & \multicolumn{1}{c}{\textless  0.001} \\ 
\multicolumn{1}{c}{Ground-based action vs HAC1H}      & -0.031 & (-0.056, -0.005) & \multicolumn{1}{c}{0.121} \\ 
\multicolumn{1}{c}{HAC1F vs HAC1H}                    & -0.008 & (-0.031, 0.015) & \multicolumn{1}{c}{1.000} \\ 
\multicolumn{1}{c}{HAC1R vs HAC1H}                    & -0.005 & (-0.032, 0.022) & \multicolumn{1}{c}{1.000} \\ 
\multicolumn{1}{c}{Ground-based action vs Air Tanker} & -0.138 & (-0.170, -0.105) & \multicolumn{1}{c}{\textless 0.001} \\ 
\multicolumn{1}{c}{HAC1F vs Air Tanker}               & -0.115 & (-0.145, -0.085) & \multicolumn{1}{c}{\textless 0.001} \\ 
\multicolumn{1}{c}{HAC1R vs Air Tanker}               & -0.112 & (-0.145, -0.079) & \multicolumn{1}{c}{\textless 0.001} \\ 
\multicolumn{1}{c}{HAC1F vs Ground-based action}      & 0.023 & (-0.009, 0.055) & \multicolumn{1}{c}{0.749} \\ 
\multicolumn{1}{c}{HAC1R vs Ground-based action}      & 0.026 & (-0.009, 0.061) & \multicolumn{1}{c}{0.749} \\ 
\multicolumn{1}{c}{HAC1R vs HAC1F}                    & 0.003 & (-0.030, 0.036) & \multicolumn{1}{c}{1.000} \\ 
\hline \\[-1.8ex] 
\end{tabular} 
		}
\begin{flushleft}
\begin{small}
\begin{textit}
All estimates are adjusted using the augmented overlap weights estimator for Initial Spread Index, Fire Weather Index, year, how the fire was discovered, ecological region, fuel type, period of day, month of the year, response time,  number of fires active, and $\ln$ of the size of the fire at the time of the initial attack. Abbreviations: HAC1H = heli-attack crew with helicopter but no rappel capability, HAC1R = heli-attack crew with helicopter and rappel capability, HAC1F = fire-attack crew with or without a helicopter and no rappel capability. P-values are adjusted for multiple comparisons using the Holm method.
\end{textit}
\end{small}
\end{flushleft}

\end{table} 

\section{Discussion}
The initial motivation for this work was to compare the effect of five initial attack interventions on wildfire growth. Multiple analytical challenges were expected in attempting to estimate this effect, including important confounding by indication, possible non-positivity and unknown model specifications. We reviewed studies evaluating the empirical performance of adjustment methods and identified those that appeared best suited to address these challenges: the overlap weights, the bias-corrected matching and the targeted maximum likelihood estimation. 

Regarding the overlap weight estimator, we demonstrated that when it is augmented with an outcome regression, it benefits from a property analogue to double-robustness in the mutli-level treatment case, following the proof of \cite{mao2019propensity} for the binary case. Additionally, we have proposed a simple asymptotic variance estimator for the overlap weights and the augmented overlap weights based on the semi-parametric theory. These variance estimators offer multiple benefits. First, they are expressed as the sample variance of a relatively simple quantity that do not require advanced coding or mathematical skills to compute. This is particularly true for the overlap weights, since estimation and inferences correspond to those produced by common generalized estimating equations routines with a robust variance estimator. A further advantage of our proposed variance estimator is that it is agnostic to the model used for computing the point estimates. Hence, it can be computed as easily whether the point estimates are obtained with parametric regression models or machine learning algorithms.

We have also conducted a simulation study comparing the overlap weights, the augmented overlap weights, the bias-corrected matching and the TMLE estimator, as well as standardization, regular inverse probability weighting and matching as benchmark comparators. Our proposed variance estimator for the overlap weight estimators performed well, even with small sample sizes. We also observed that methods that combine outcome and treatment modeling (bias-corrected matching, TMLE and augmented overlap weights) performed better than those solely based on the treatment model (inverse probability weighting, matching and overlap weights) in terms of bias reduction, standard deviation of estimates and RMSE. Under positivity violations, inverse probability weighting and matching yielded biased estimates with increased variability, while all other methods were relatively unaffected. We also observed that the variance estimator for the matching and bias-corrected matching estimators of \cite{scotina2019matching} performed poorly in the plasmode simulation. We hypothesize that this is because of the binary nature of the outcome.

Unsurprisingly, when an incorrect parametric implementation was used, all adjustment methods produced biased estimates. However, the bias for matching and bias-corrected matching tended to get smaller as sample size increased, whereas the bias remained constant for the other methods. A plausible explanation for this phenomenon is that these matching methods are less model dependent, since they involve imputing the missing counterfactual outcomes by the observed outcomes from subjects in the other treatment groups. As sample size increases, the pool of control subjects get larger, thus allowing observations to be matched with others that are more similar to them. We did not observe that methods that combine outcome and exposure modeling performed better than the others under models misspecification. While it is expected that double-robustness theoretically helps protecting against the bias attributable to model misspecifications, this property requires that at least one of the two models involved is correct for producing unbiased estimates. When both models are incorrect to some extent, as should arguably be expected in practice, double-robust methods do not necessarily produce estimates with less bias than others \citep{kreif2016evaluating, kang2007demystifying}. 

The results of our simulation study finally indicate that machine learning methods can be helpful for preventing misspecification bias for some, but not all, adjustment methods. Indeed, for all methods that include an outcome modeling component (standardization, bias-corrected matching, TMLE and augmented overlap weight), the results under a machine learning implementations were very similar to those obtained under the ideal case of a correct parametric specification for all considered performance metrics. As such, the routine use of machine learning algorithms for performing confounders adjustment may have benefits to help preventing bias without paying a price in terms of power. On the other hand, methods that focus only on modeling the treatment (inverse probability weighting, matching and overlap weight) had considerable bias when employing a machine learning implementation. A possible explanation for these results is that machine learning algorithms may tend to more often predict treatment probabilities close to 0 or 1 because they are better able to fit the observed data, thus exacerbating practical positivity violations. 

Standardization was the method that overall performed best, closely followed by the augmented overlap weight in most scenarios and sometimes also by TMLE. In practice, we warn that the choice of an approach should not only be guided by its performance, but also by the question of interest. For example, if the goal is to inform about a decision that would apply at the whole population level, the average effect in the entire population would be the one that is of most interest, whereas if the question is to help the decision in cases where several treatment options are possible, the methods that estimate an effect in the subgroups where there is overlap would likely be more appropriate. 

Based on the context of the wildfire growth problem and the results of the simulation study, we decided to estimate the effect of initial attack interventions using the augmented overlap weight estimator, implemented with machine learning algorithms. The adjusted associations we observed were counter-intuitive: the more aggressive the intervention was, the larger was the probability that fires grew. These associations are very unlikely to represent the true causal effect. Based on our simulation study's results, we do not expect these results can be explained by residual confounding due to measured confounders or to non-positivity. We thus hypothesize that residual confounding due to important non-measured confounders is present, despite the fact we have made important efforts to include multiple potential confounders in our analysis. Having accounted for the factors previously shown to affect the probabilities of fire growth after initial attack in this system \citep{arienti2006empirical}, we are unable to advance a more precise explanation.

Ultimately, the findings from this study originate from simulation studies and are thus limited to the contexts that were investigated. However, the plasmode simulation based on real data helped understanding the performance of these approaches in a realistic setting. Bias elimination relies on the fact that all confounders are measured, which arguably never occurs in practice. As such, at least some amount of residual confounding due to unmeasured confounders is inevitably present in real data analyses. This is most likely the explanation for the counter-intuitive associations we have observed in the wildfire analysis. While comparing the effect of fire-fighting interventions using observational data is fraught with multiple challenges, we believe it is essential to better inform fire managers regarding the best course of action when wildfires occur. We hope our study will stimulate others to attempt facing these challenges. 

\quad

\textbf{Acknowledgments} 

\quad

This research was funded by the Natural Sciences and Engineering Research Council of Canada grant \# 2016-06295. S. A. Diop was also supported by a scholarship of Excellence from the Facult\'{e} de sciences et de g\'{e}nie of Universit\'{e} Laval. D Talbot is a Fonds de recherche du Qu\'{e}bec - Sant\'{e} Chercheur-Boursier. 


\newpage

\clearpage 
\bibliography{myrefs}

\begin{thebibliography}{38}
\providecommand{\natexlab}[1]{#1}
\providecommand{\url}[1]{\texttt{#1}}
\expandafter\ifx\csname urlstyle\endcsname\relax
  \providecommand{\doi}[1]{doi: #1}\else
  \providecommand{\doi}{doi: \begingroup \urlstyle{rm}\Url}\fi

\bibitem[Abadie and Imbens(2006)]{abadie2006large}
Alberto Abadie and Guido~W Imbens.
\newblock Large sample properties of matching estimators for average treatment
  effects.
\newblock \emph{econometrica}, 74\penalty0 (1):\penalty0 235--267, 2006.

\bibitem[Abadie and Imbens(2011)]{abadie2011bias}
Alberto Abadie and Guido~W Imbens.
\newblock Bias-corrected matching estimators for average treatment effects.
\newblock \emph{Journal of Business \& Economic Statistics}, 29\penalty0
  (1):\penalty0 1--11, 2011.

\bibitem[Agriculture and Forestry(2015)]{historicalwildfiredatabase}
Wildfire Management Branch~Alberta Agriculture and Forestry.
\newblock Historical wildfire database.
\newblock \emph{visited on Monday 25 October 2018 03:57:00 PM}, 2015.
\newblock URL
  \url{http://wildfire.alberta.ca/resources/historical-data/historical-wildfire-database.aspx}.

\bibitem[Arienti et~al.(2006)Arienti, Cumming, and
  Boutin]{arienti2006empirical}
M~Cecilia Arienti, Steven~G Cumming, and Stan Boutin.
\newblock Empirical models of forest fire initial attack success probabilities:
  the effects of fuels, anthropogenic linear features, fire weather, and
  management.
\newblock \emph{Canadian Journal of Forest Research}, 36\penalty0
  (12):\penalty0 3155--3166, 2006.

\bibitem[Austin(2008)]{austin2008performance}
Peter~C Austin.
\newblock The performance of different propensity-score methods for estimating
  relative risks.
\newblock \emph{Journal of clinical epidemiology}, 61\penalty0 (6):\penalty0
  537--545, 2008.

\bibitem[Austin(2010)]{austin2010performance}
Peter~C Austin.
\newblock The performance of different propensity-score methods for estimating
  differences in proportions (risk differences or absolute risk reductions) in
  observational studies.
\newblock \emph{Statistics in medicine}, 29\penalty0 (20):\penalty0 2137--2148,
  2010.

\bibitem[Austin et~al.(2007)Austin, Grootendorst, and
  Anderson]{austin2007comparison}
Peter~C Austin, Paul Grootendorst, and Geoffrey~M Anderson.
\newblock {A comparison of the ability of different propensity score models to
  balance measured variables between treated and untreated subjects: a Monte
  Carlo study}.
\newblock \emph{Statistics in Medicine}, 26\penalty0 (4):\penalty0 734--753,
  2007.

\bibitem[Cumming(2005)]{cumming2005effective}
SG~Cumming.
\newblock Effective fire suppression in boreal forests.
\newblock \emph{Canadian Journal of Forest Research}, 35\penalty0 (4):\penalty0
  772--786, 2005.

\bibitem[Feng et~al.(2012)Feng, Zhou, Zou, Fan, and Li]{feng2012generalized}
Ping Feng, Xiao-Hua Zhou, Qing-Ming Zou, Ming-Yu Fan, and Xiao-Song Li.
\newblock Generalized propensity score for estimating the average treatment
  effect of multiple treatments.
\newblock \emph{Statistics in medicine}, 31\penalty0 (7):\penalty0 681--697,
  2012.

\bibitem[Fr{\"o}lich(2004)]{frolich2004programme}
Markus Fr{\"o}lich.
\newblock Programme evaluation with multiple treatments.
\newblock \emph{Journal of Economic Surveys}, 18\penalty0 (2):\penalty0
  181--224, 2004.

\bibitem[Halekoh et~al.(2006)Halekoh, H{\o}jsgaard, Yan, et~al.]{halekoh2006r}
Ulrich Halekoh, S{\o}ren H{\o}jsgaard, Jun Yan, et~al.
\newblock The r package geepack for generalized estimating equations.
\newblock \emph{Journal of Statistical Software}, 15\penalty0 (2):\penalty0
  1--11, 2006.

\bibitem[Hern{\'a}n and Robins(2020)]{hernan2017causal}
M.A. Hern{\'a}n and J.M. Robins.
\newblock \emph{Causal Inference: What if}.
\newblock Boca Raton: Chapman \& Hall/CRC. 2020.

\bibitem[Holm(1979)]{holm1979simple}
Sture Holm.
\newblock A simple sequentially rejective multiple test procedure.
\newblock \emph{Scandinavian journal of statistics}, pages 65--70, 1979.

\bibitem[Horvitz and Thompson(1952)]{horvitz1952generalization}
Daniel~G Horvitz and Donovan~J Thompson.
\newblock A generalization of sampling without replacement from a finite
  universe.
\newblock \emph{Journal of the American statistical Association}, 47\penalty0
  (260):\penalty0 663--685, 1952.

\bibitem[Huber et~al.(2013)Huber, Lechner, and Wunsch]{huber2013performance}
Martin Huber, Michael Lechner, and Conny Wunsch.
\newblock The performance of estimators based on the propensity score.
\newblock \emph{Journal of Econometrics}, 175\penalty0 (1):\penalty0 1--21,
  2013.

\bibitem[Kang and Schafer(2007)]{kang2007demystifying}
Joseph~DY Kang and Joseph~L Schafer.
\newblock Demystifying double robustness: A comparison of alternative
  strategies for estimating a population mean from incomplete data.
\newblock \emph{Statistical science}, 22\penalty0 (4):\penalty0 523--539, 2007.

\bibitem[Kooperberg(2019)]{Kooperberg2019polspline}
Charles Kooperberg.
\newblock \emph{polspline: Polynomial Spline Routines}, 2019.
\newblock URL \url{https://CRAN.R-project.org/package=polspline}.
\newblock R package version 1.1.14.

\bibitem[Kooperberg et~al.(1997)Kooperberg, Bose, and
  Stone]{koop1997polychotomous}
Charles Kooperberg, Smarajit Bose, and Charles~J Stone.
\newblock Polychotomous regression.
\newblock \emph{Journal of the American Statistical Association}, 92\penalty0
  (437):\penalty0 117--127, 1997.

\bibitem[Kreif et~al.(2016)Kreif, Gruber, Radice, Grieve, and
  Sekhon]{kreif2016evaluating}
No{\'e}mi Kreif, Susan Gruber, Rosalba Radice, Richard Grieve, and Jasjeet~S
  Sekhon.
\newblock Evaluating treatment effectiveness under model misspecification: a
  comparison of targeted maximum likelihood estimation with bias-corrected
  matching.
\newblock \emph{Statistical methods in medical research}, 25\penalty0
  (5):\penalty0 2315--2336, 2016.

\bibitem[Li and Li(2019)]{li2018propensity}
Fan Li and Fan Li.
\newblock Propensity score weighting for causal inference with multi-valued
  treatments.
\newblock \emph{Annals of Applied Statistics}, 2019.

\bibitem[Li et~al.(2018)Li, Morgan, and Zaslavsky]{li2018balancing}
Fan Li, Kari~Lock Morgan, and Alan~M Zaslavsky.
\newblock Balancing covariates via propensity score weighting.
\newblock \emph{Journal of the American Statistical Association}, 113\penalty0
  (521):\penalty0 390--400, 2018.

\bibitem[Li and Greene(2013)]{li2013weighting}
Liang Li and Tom Greene.
\newblock A weighting analogue to pair matching in propensity score analysis.
\newblock \emph{The international journal of biostatistics}, 9\penalty0
  (2):\penalty0 215--234, 2013.

\bibitem[Liaw and Wiener(2002)]{Liaw2002Classification}
Andy Liaw and Matthew Wiener.
\newblock Classification and regression by randomforest.
\newblock \emph{R News}, 2\penalty0 (3):\penalty0 18--22, 2002.
\newblock URL \url{https://CRAN.R-project.org/doc/Rnews/}.

\bibitem[Lunceford and Davidian(2004)]{lunceford2004stratification}
Jared~K Lunceford and Marie Davidian.
\newblock Stratification and weighting via the propensity score in estimation
  of causal treatment effects: a comparative study.
\newblock \emph{Statistics in medicine}, 23\penalty0 (19):\penalty0 2937--2960,
  2004.

\bibitem[Luque-Fernandez et~al.(2018)Luque-Fernandez, Schomaker, Rachet, and
  Schnitzer]{luque2018targeted}
Miguel~Angel Luque-Fernandez, Michael Schomaker, Bernard Rachet, and Mireille~E
  Schnitzer.
\newblock Targeted maximum likelihood estimation for a binary treatment: A
  tutorial.
\newblock \emph{Statistics in medicine}, 37\penalty0 (16):\penalty0 2530--2546,
  2018.

\bibitem[Mao et~al.(2019)Mao, Li, and Greene]{mao2019propensity}
Huzhang Mao, Liang Li, and Tom Greene.
\newblock Propensity score weighting analysis and treatment effect discovery.
\newblock \emph{Statistical methods in medical research}, 28\penalty0
  (8):\penalty0 2439--2454, 2019.

\bibitem[Polley et~al.(2018)Polley, LeDell, Kennedy, and {van der
  Laan}]{Polley2018SuperLearner}
Eric Polley, Erin LeDell, Chris Kennedy, and Mark {van der Laan}.
\newblock \emph{SuperLearner: Super Learner Prediction}, 2018.
\newblock URL \url{https://CRAN.R-project.org/package=SuperLearner}.
\newblock R package version 2.0-24.

\bibitem[Robins(1997)]{Robins1997}
James~M Robins.
\newblock Marginal structural models.
\newblock \emph{Proceedings of the Section on Bayesian Statistical Science.},
  1997.
\newblock American Statistical Association: Alexandria, VA, 1998; 1-10.

\bibitem[Robins(2000)]{robins2000marginalb}
James~M Robins.
\newblock Marginal structural models versus structural nested models as tools
  for causal inference.
\newblock In \emph{Statistical models in epidemiology, the environment, and
  clinical trials}, pages 95--133. Springer, 2000.

\bibitem[Scotina et~al.(2019)Scotina, Beaudoin, and
  Gutman]{scotina2019matching}
Anthony~D Scotina, Francesca~L Beaudoin, and Roee Gutman.
\newblock Matching estimators for causal effects of multiple treatments.
\newblock \emph{Statistical methods in medical research}, page
  0962280219850858, 2019.

\bibitem[Siddique et~al.(2019)Siddique, Schnitzer, Bahamyirou, Wang, Holtz,
  Migliori, Sotgiu, Gandhi, Vargas, Menzies, et~al.]{siddique2019causal}
Arman~Alam Siddique, Mireille~E Schnitzer, Asma Bahamyirou, Guanbo Wang,
  Timothy~H Holtz, Giovanni~B Migliori, Giovanni Sotgiu, Neel~R Gandhi, Mario~H
  Vargas, Dick Menzies, et~al.
\newblock Causal inference with multiple concurrent medications: A comparison
  of methods and an application in multidrug-resistant tuberculosis.
\newblock \emph{Statistical methods in medical research}, 28\penalty0
  (12):\penalty0 3534--3549, 2019.

\bibitem[Tremblay et~al.(2018)Tremblay, Duchesne, and
  Cumming]{tremblay2018survival}
Pier-Olivier Tremblay, Thierry Duchesne, and Steven~G Cumming.
\newblock Survival analysis and classification methods for forest fire size.
\newblock \emph{PLoS ONE}, 13\penalty0 (1):\penalty0 e0189860, 2018.

\bibitem[Van~der Laan and Rose(2011)]{van2011targeted}
Mark~J Van~der Laan and Sherri Rose.
\newblock \emph{Targeted learning: causal inference for observational and
  experimental data}.
\newblock Springer Science \& Business Media, 2011.

\bibitem[Van Der~Laan and Rubin(2006)]{van2006targeted}
Mark~J Van Der~Laan and Daniel Rubin.
\newblock Targeted maximum likelihood learning.
\newblock \emph{The International Journal of Biostatistics}, 2\penalty0 (1),
  2006.

\bibitem[Van~der Laan et~al.(2007)Van~der Laan, Polley, and
  Hubbard]{van2007super}
Mark~J Van~der Laan, Eric~C Polley, and Alan~E Hubbard.
\newblock Super learner.
\newblock \emph{Statistical applications in genetics and molecular biology},
  6\penalty0 (1), 2007.

\bibitem[Yang et~al.(2016)Yang, Imbens, Cui, Faries, and
  Kadziola]{yang2016propensity}
Shu Yang, Guido~W Imbens, Zhanglin Cui, Douglas~E Faries, and Zbigniew
  Kadziola.
\newblock Propensity score matching and subclassification in observational
  studies with multi-level treatments.
\newblock \emph{Biometrics}, 72\penalty0 (4):\penalty0 1055--1065, 2016.

\bibitem[Yoshida et~al.(2017)Yoshida, Hern{\'a}ndez-D{\'\i}az, Solomon,
  Jackson, Gagne, Glynn, and Franklin]{yoshida2017matching}
Kazuki Yoshida, Sonia Hern{\'a}ndez-D{\'\i}az, Daniel~H Solomon, John~W
  Jackson, Joshua~J Gagne, Robert~J Glynn, and Jessica~M Franklin.
\newblock Matching weights to simultaneously compare three treatment groups:
  comparison to three-way matching.
\newblock \emph{Epidemiology}, 28\penalty0 (3):\penalty0 387, 2017.

\bibitem[Zou(2009)]{zou2009assessment}
GY~Zou.
\newblock Assessment of risks by predicting counterfactuals.
\newblock \emph{Statistics in medicine}, 28\penalty0 (30):\penalty0 3761--3781,
  2009.

\end{thebibliography}
\end{document}